\documentclass[preprint,1p,number]{elsarticle}
\usepackage[usenames,dvipsnames]{color}
\usepackage{graphicx}
\usepackage{multirow}

\newcommand{\secref}[1]{Section~\ref{sec:#1}}

\newcommand{\figref}[1]{Fig.~\ref{fig:#1}}
\newcommand{\tblref}[1]{Table~\ref{tbl:#1}}
\newcommand{\eqref}[1]{Eq.~(\ref{eq:#1})}

\journal{Physica A}

\begin{document}

\begin{frontmatter}

\title{Self-similar scaling of density in complex real-world networks}
\author{Neli Blagus\corref{coraut}}
\ead{neli.blagus@fri.uni-lj.si}
\author{Lovro \v Subelj}
\ead{lovro.subelj@fri.uni-lj.si}
\author{Marko Bajec}
\ead{marko.bajec@fri.uni-lj.si}
\address{University of Ljubljana, Faculty of Computer and Information Science, Ljubljana, Slovenia}
\cortext[coraut]{Corresponding author. Tel.: +386 1 476 81 86.}


\begin{abstract}
Despite their diverse origin, networks of large real-world systems reveal a number of common properties including small-world phenomena, scale-free degree distributions and modularity. Recently, network self-similarity as a natural outcome of the evolution of real-world systems has also attracted much attention within the physics literature. Here we investigate the scaling of density in complex networks under two classical box-covering renormalizations---network coarse-graining---and also different community-based renormalizations. The analysis on over $50$ real-world networks reveals a power-law scaling of network density and size under adequate renormalization technique, yet irrespective of network type and origin. The results thus advance a recent discovery of a universal scaling of density among different real-world networks [Laurienti~et~al., Physica~A 390~(20) (2011) 3608-3613.] and imply an existence of a scale-free density also within---among different self-similar scales of---complex real-world networks. The latter further improves the comprehension of self-similar structure in large real-world networks with several possible applications.
\end{abstract}
 

\begin{keyword}
complex networks \sep self-similarity \sep network density \sep community structure
\\
\textit{PACS:} 89.75.Hc \sep 05.45.Df \sep 89.75.Da \sep 89.75.Fb
\end{keyword}

\end{frontmatter}


\section{\label{sec:intro}Introduction}
The study of complex real-world networks and underlying systems has erupted in recent years in various fields of science. Due to their simple and intelligible form, networks enable representation of diverse systems of complex interactions and provide for their common investigation. Thus, several fundamental properties of large real-world networks have been revealed in the past decade. These include small-world phenomena~\cite{WS98}, scale-free degree distributions~\cite{FFF99,BA99}, network clustering~\cite{WS98,SV05} and robustness~\cite{AJB00,CEBH00}, degree mixing~\cite{New02,New03b}, community and hierarchical structure~\cite{GN02,RB03}, network motifs~\cite{MSIKCA01} and other~\cite{ABL10} (for reviews see~\cite{New08,New03a}). More recently, network self-similarity as an inherent property behind the evolution of real-world systems has also attracted much of attention within the physics community~\cite{GDDGA03,SHM05,ILKMIA05,KGKK07a,RRBF08,GSM07}. 

\begin{figure}[tb]
\centering
\includegraphics[width=0.75\columnwidth]{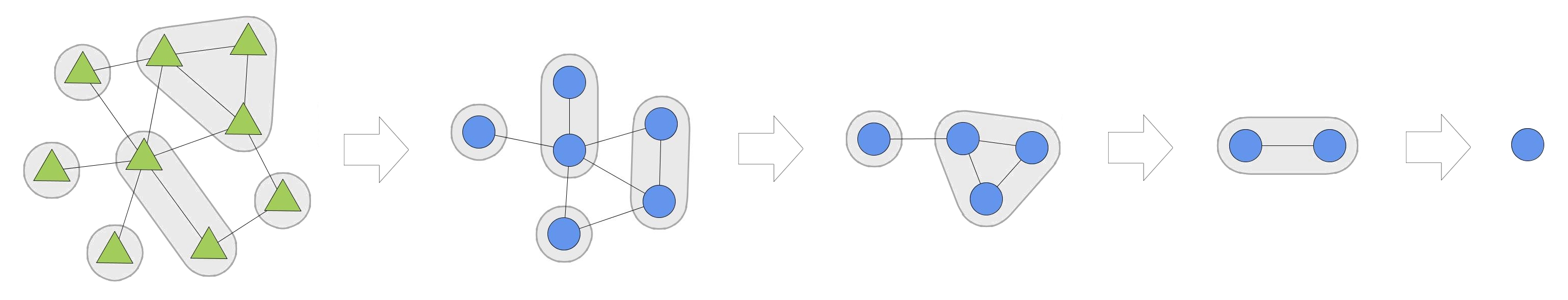}
\caption{\label{fig:renorm}Network renormalization---system coarse-graining---technique~\cite{SHM05,GSKK06} applied to a small example network. At each step, the network is covered with boxes that are replaced by super-nodes. The latter are linked when a corresponding link also exists in the (original) network. The process then repeats until only a single node remains or multiple nodes in the case of a disconnected network. (Here the network is randomly tiled with boxes of nodes at a distance smaller than $2$.)}
\end{figure}

Network self-similarity is commonly considered alongside the concept of fractal networks~\cite{SHM05,CR10}. Fractality is a property of a geometric object that it is exactly or approximately similar to a part of itself~\cite{Man83}. Nevertheless, classical theory of self-similarity requires a power-law scaling between the system size and its parts under some renormalization~\cite{BH96,BH00}. The latter is an iterative process where a system is coarse-grained into smaller replicas, thus its essential structural features are preserved~\cite{SHM05,GSKK06} (\figref{renorm}). Hence, fractal or self-similar networks commonly refer only to a self-similar scaling exponent in the aforementioned power-law relation~\cite{SHM05,SHM06,KGKK07a,RGSM08}. However, network self-similarity is also investigated in the context of other network properties~\cite{GDDGA03,ILKMIA05,KGKK07a,RRBF08,BRV01} under various renormalization techniques~\cite{SHM05,GSKK06,SGHM07,BSPG11}.\footnote{Throughout the paper we refer to network self-similarity in a general sense.} (Note that fractal scaling laws observed in real-world networks do not necessarily imply a self-similar network~\cite{SBCGP11}.)

Guimer\`{a}~et~al.~\cite{GDDGA03} have first observed self-similar community size distributions in a network of human communications. Furthermore, Song~et~al.~\cite{SHM05,SHM06} have proposed an adequate renormalization technique (\figref{renorm}) to expose the origin of self-similar fractal scaling in web, collaboration and different biological networks. The latter in fact gives rise to degree disassortativity~\cite{SHM06} and resilience to diseases~\cite{ZZZC08}, commonly observed for these networks. Still, such scaling cannot coexist with a small-world network topology~\cite{CS04,KY10}. Self-similarity has also been considered as a scale-invariance of degree distribution~\cite{KGKK07a,RGSM08} or maximum degree~\cite{RRBF08,RBFR09} under network renormalization, while Itzkovitz~et~al.~\cite{ILKMIA05} have revealed self-dissimilarity in a motif structure for different biological and technological networks. Authors have also considered network self-similarity in the context of different dynamical processes including percolation~\cite{SKB11} and synchronization~\cite{ZL11}.

Despite the above efforts, there is yet little evidence whether self-similarity exists only in certain networks and which properties are indeed invariant throughout different network scales. We thus here investigate the scaling of density---defined as the number of links to all possible links---with respect to network size under five renormalization techniques borrowed from the field of fractal networks~\cite{SHM05,SHM06} and community detection literature~\cite{For10,POM09}. Analysis on over $50$ real-world networks of diverse origin reveals a self-similar power-law scaling of network density and size (under suitable renormalization). The latter advances a recent work of~Laurienti~et~al.~\cite{LJTBH11} who have observed a universal scaling of density among different real-world networks, while Leskovec~et~al.~\cite{LKF05,LKF07} have also found similar densification laws in evolving networks. The results thus imply an existence of a scale-free density not only among, but also within---among different self-similar scales of---complex networks irrespective of their type and the underlying domain. Hence, under adequate renormalization self-similar real-world networks neither get denser nor sparser with respect to their size, whereas characteristic network topology is also largely retained throughout the renormalization.

The rest of the paper is structured as follows. \secref{prelim} introduces different renormalization techniques and real-world network data adopted in the research. Empirical analysis with formal discussion on real-world and random networks is presented in~\secref{analys}, while~\secref{conc} gives final conclusions and discusses future work.


\section{\label{sec:prelim}Techniques and network data}
Self-similarity is primarily studied under the framework of network renormalization~\cite{SHM05,KGKK07a}. As already discussed, renormalization is an iterative coarse-graining technique, where the original network is covered with boxes, thus each node belongs to exactly one box~\cite{SHM05,GSKK06} (\figref{renorm}). Boxes are then replaced by super-nodes that are linked when a corresponding link also exists in the (original) network. The entire process repeats until no links remain and the number of nodes equals to the number of connected components. 

While there exists a number of different box-covering approaches, not all of them are able to reveal self-similar scales in complex networks. Thus, we employ techniques that have already proven useful for exposing self-similarity in various real-world networks~\cite{GDDGA03,SHM05,KGKK07a,RRBF08}. In particular, we adopt methods commonly used in analysis of fractal networks, as well as different community detection algorithms.

Fractal network structure is mainly explored under two general classes of renormalization techniques, namely, node coloring and network burning approaches~\cite{SHM05,SGHM07} (for reviews see~\cite{GSM07, RGSM08}). In the former, box-covering is mapped to a node coloring problem~\cite{Chr71,Wil84}, whereas, in the latter, boxes are grown around a randomly selected seed node. Although there exist several efficient algorithms for node coloring~\cite{Wil84,CLRS01}, network burning methods offer some distinct advantages~\cite{SGHM07}. Different authors have proposed a wide range of alternative network coarse-graining techniques including methods based on connectivity patterns~\cite{ILKMIA05}, skeleton of the network~\cite{GSKK06}, link-covering~\cite{ZJS07} and other~\cite{KGKK07a,KGKK07b,BSPG11,ZL11}. 

For the purpose of this research, we adopt two classical network burning approaches. First, box-tiling method, randomly tiles the network with boxes of nodes that are at a distance smaller than~$l_B$~\cite{SHM05,SHM06} (\figref{renorm}). Second, cluster-growing method, incrementally grows boxes from randomly selected seed nodes within a distance not larger than~$r_B$~\cite{SGHM07,KGKK07b}. Hence, for random configurations, $l_B=2\cdot r_B+1$~\cite{SGHM07}. Box-tilling method allows for somewhat easier analytical consideration, whereas cluster-growing approach enables more efficient implementation. For the analysis in~\secref{analys}, we set $l_B$ to $3$ and $r_B$ to $2$ with respect to network small-worlds~\cite{WS98}. Note that the latter extends the definition of an egonet~\cite{Fre82,EB05}---a subnetwork inferred by a central ego node and its neighbors---which can be seen as a local signature of the respective~node.

\begin{table}[p]
\centering
\caption{\label{tbl:nets}Real-world networks. ($n$ and $m$ correspond to the number of nodes and links, respectively.)}
\scriptsize
\begin{tabular}{ccrr}
\\\hline\noalign{\smallskip}
Network & Type & \multicolumn{1}{c}{$n$} & \multicolumn{1}{c}{$m$}\\\noalign{\smallskip}\hline\noalign{\smallskip}
Zachary's karate club~\cite{zachary1977information} & \multirow{3}{*}{Social} & $34$ & $78$ \\ 
Lusseau's dolphins~\cite{lusseau2003bottlenose} & & $62$ & $159$ \\
Comp. sci. PhD students~\cite{BM-D} & & $1025$ & $1043$ \\ 
\hline\noalign{\smallskip}
\textit{Facebook} friendships~\cite{Sub-D} & \multirow{2}{*}{On-line social} & $324$ & $2218$ \\ 
\textit{Wikipedia} who-votes-who~\cite{LHK10} & & $7066$ & $100736$ \\ 
\hline\noalign{\smallskip}
Slovenian comp. science~\cite{Sub-D}  & \multirow{8}{*}{Collaboration} & $239$ & $568$ \\ 
Krebs's Internet industry~\cite{BM-D} & & $219$ & $630$ \\ 
Complex networks science~\cite{New06} & & $379$ & $914$ \\
Paul Erd\H{o}s collaborations~\cite{BM-D} & & $446$ & $1413$ \\ 
\textit{Comput. Geometry} archive~\cite{jonescomputational} & & $3621$ & $9461$ \\
\textit{General Relativity} archive~\cite{LKF05} & & $4158$ & $13422$ \\
\textit{PGP} web-of-trust~\cite{boguna2004models} & & $10680$ & $24316$ \\ 
\textit{Astro Physics} archive~\cite{LKF07} & & $17903$ & $196972$ \\
\hline\noalign{\smallskip}
US political books~\cite{adamic2005political}  & Co-purchase  & $105$ & $441$ \\ 
\hline\noalign{\smallskip}
\textit{amazon.com} domain~\cite{subelj2011ubiquitousness}  & \multirow{4}{*}{Web graph} & $2879$ & $3886$\\
\textit{epa.gov} domain~\cite{BM-D}  & & $4253$ & $8897$ \\
Broad-topic queries~\cite{kleinberg1999authoritative}  & & $5925$ & $15770$ \\
US political blogs~\cite{adamic2005political}  & & $1222$ & $16714$ \\
\hline\noalign{\smallskip}
\textit{Graph Drawing} proceedings~\cite{BM-D}  & \multirow{8}{*}{Citation} & $249$ & $635$ \\ 
Stanley Milgram citations~\cite{BM-D} & & $233$ & $994$ \\
H. Small \& B. Griffith citations~\cite{BM-D} & & $1024$ & $4916$ \\ 
\textit{Scientometrics} archive~\cite{BM-D}  & & $2678$ & $10368$ \\
Teuvo Kohonen citations~\cite{BM-D}  & & $3704$ & $12673$ \\
Joshua Lederberg citations~\cite{BM-D} & & $8212$ & $41430$ \\ 
Ahmed Zewail citations~\cite{BM-D} & & $6640$ & $54173$ \\
\textit{High E. Particle Phys.} archive~\cite{hep} & & $27400$ & $352021$ \\
\hline\noalign{\smallskip}
Mobile phone records~\cite{vastChallenge} & \multirow{3}{*}{Communication} & $345$ & $355$ \\
Emails at a university~\cite{GDDGA03} & & $1133$ & $5451$ \\
Emails at \textit{Enron}~\cite{LLDM09} & & $33696$ & $180811$ \\
\hline\noalign{\smallskip}
Novel \textit{David Copperfield}~\cite{New06}  & \multirow{7}{*}{Information} & $112$ & $425$ \\
\textit{Roget's Thesaurus} dictionary~\cite{knuth1993stanford} & & $994$ & $3640$ \\	
Java documentation (\texttt{javax})~\cite{subelj2011unfolding} & & $1031$ & $4408$ \\ 
\textit{ODLIS} dictionary~\cite{odlis} & & $2898$ & $16376$ \\
\textit{USF} association norms~\cite{associations} & & $10617$ & $63782$ \\
\textit{FOLDOC} dictionary~\cite{batagelj2002network} & & $13356$ & $91471$ \\
\textit{WordNet} dictionary~\cite{BM-D} & & $75606$ & $119564$ \\  
\hline\noalign{\smallskip}
Small software project~\cite{BM-D} & \multirow{4}{*}{Software} & $83$ & $125$ \\
\textit{JUNG} graph framework~\cite{subelj2011community} & & $398$ & $943$ \\ 
Java language (\texttt{javax})~\cite{subelj2011community} & & $1570$ & $7194$ \\ 
Java language (general)~\cite{BM-D} & & $1538$ & $7817$ \\ 
\hline\noalign{\smallskip}
\textit{Oregon} aut. systems~\cite{New-D} & \multirow{2}{*}{Internet} & $22963$ & $48436$ \\
\textit{Gnutella} file sharing~\cite{LKF07} & & $36646$ & $88303$ \\
\hline\noalign{\smallskip}
European roads~\cite{SB11a} & \multirow{4}{*}{Technological} & $1039$ & $1305$ \\
Finite automaton~\cite{BM-D} & & $1096$ & $1677$ \\ 
US air lines~\cite{BM-D} & & $332$ & $2126$ \\ 
US power grid~\cite{BM-D} & & $4941$ & $6594$ \\
\hline\noalign{\smallskip}
\textit{Escherichia Coli} regulatory~\cite{BM-D} & \multirow{3}{*}{Biological} & $328$ & $456$ \\
\textit{Caenorhabditis Elegans} neural~\cite{WS98} & & $297$ & $2148$ \\ 
Yeast protein interactions~\cite{bu2003topological} & & $2224$ & $6609$ \\	
\hline\noalign{\smallskip}
Data modeling~\cite{BM-D} & Other & $638$ & $1020$\\
\hline\noalign{\smallskip}
\textit{Amazon} products~\cite{LAH07} & Co-purchase  & $524366$ & $1491774$ \\\noalign{\smallskip}
\textit{nd.edu} domain~\cite{AJB99}  & Web graph & $325729$ & $1497135$\\\noalign{\smallskip}
Pennsylvania roads~\cite{LLDM09} & Technological & $1087562$ & $1541514$ \\\noalign{\smallskip}
\textit{Wikipedia} talk service~\cite{LHK10} & Communication & $2388953$ & $4656682$ \\\noalign{\smallskip}
\textit{Skitter} overlay map~\cite{LKF05} & Internet & $1694616$ & $11094209$\\\noalign{\smallskip}\hline
\end{tabular}
\end{table}

We further adopt several algorithms drawn from community detection literature (for reviews see~\cite{For10,POM09}). Here boxes are identified by communities~\cite{GN02}---groups of nodes densely connected within and only loosely connected between---revealed with selected algorithm, whereas network coarse-graining procedure is else identical as above. Community detection has already been successfully employed to reveal self-similarity in real-world networks~\cite{GDDGA03}. Recent work also implies an existence of community structures on various scales of complex real-world networks~\cite{TKKKKS11,HAMND11}. Hence, community detection appears to be an adequate alternative to classical box-covering renormalization techniques.

Due to generality, we consider three diverse community detection algorithms. First, we adopt balanced propagation~\cite{SB11a} as an example of a highly scalable state-of-the-art algorithm. The approach is based on the label propagation principle of Raghavan~et~al.~\cite{RAK07}, while node balancers are introduced to improve the stability of the algorithm (stability parameter is set to $1/4$). Next, we employ a fast hierarchical optimization of modularity~$Q$~\cite{NG04} proposed by Clauset~et~al.~\cite{CNM04} as one of most widely used approaches in the past literature~\cite{For10}. However, due to many limitations of the measure of modularity~$Q$, high values of~$Q$ cannot be regarded as an indication of network community structure~\cite{FB07,GDC10}. Last, we also consider a spectral algorithm of Newman~\cite{New06} as a representative of a partitioning approach with origins in classical graph theory~\cite{Bol98}. The algorithm reveals communities by extracting the leading eigenvector of network modularity matrix using a power method. 

Analysis in~\secref{analys} is conducted on $55$ real-world networks that are often analyzed in complex network literature (\tblref{nets}), and also on random graphs \'{a}~la~Erd\H{o}s-R{\'e}nyi~\cite{ER59} and different generative graph models. The real-world networks range between tens of nodes and tens of millions of links; and include different social---classical, on-line, collaboration etc.; information---web graphs, citation, communication etc.; technological---Internet, software, transportation etc.; biological---protein, genetic and neural; and other networks. Due to the large number of networks considered, detailed description is omitted. Still, networks were carefully chosen thus to represent a relatively diverse set of real-world systems including most types of networks commonly analyzed in the literature. For simplicity, all networks are considered as simple undirected graphs and reduced to largest connected components.


\section{\label{sec:analys}Analysis and discussion}
In the following we first analyze self-similar scaling of density in real-world networks of moderate size (\secref{analys:real}), while analysis on Erd\H{o}s-R{\'e}nyi random graphs and different generative graph models is given in \secref{analys:rand}. In~\secref{analys:large} we further consider self-similarity of five larger real-world networks with at least a million links.


\subsection{\label{sec:analys:real}Real-world networks}

The algorithms were first applied to 50 real-world networks (\tblref{nets}). According to the number of nodes $n$ and density $d$ from original and reduced networks we examine the density scaling with respect to network size. In particular, $d$ is expressed as a power function of $n$ through formula $d = c \cdot n^{-\gamma}$, where $\gamma$ is a scaling exponent and $c$ is a constant. We measure goodness of fit to the data using coefficient of determination $R^2$---how well the network size predicts density---and dependence between both variables corresponding to Spearman's correlation coefficient $\rho$---the extent to which network density decreases as network size increases. Moreover, we also evaluate the number of self-similar scales $S$ defining how many renormalized networks are revealed under different techniques.

Mean estimates for each method appear in \tblref{real}. Coefficients $R^2$ demonstrate that the power-law relationship between the size and density appears to be a good fit to the data under box-covering methods and balanced propagation based renormalization. (We can reject the null hypothesis---no actual relationship between variables---at one percent significance level, thus results are statistically significant.) Irrespective of renormalization technique, $R^2$ and $\rho$ for original networks are improved considering also their renormalized varieties. Otherwise, box-covering methods perform better than community detection algorithms, whereas balanced propagation exhibits the most homogeneous relationship between size and density. Spectral algorithm and modularity optimization prove the worst, particularly at observing fits for renormalized networks only. In the case of modularity optimization, this could be largely due to its resolution limit~\cite{FB07}, lack of global maximum and degeneracy of optimal partitions~\cite{GDC10}. On the other hand, spectral analysis is in fact an optimization of eigenvectors of the modularity matrix. Therefore, it is attributed to the above mentioned modularity limitations, whereas it also reveals modules in random networks~\cite{LF09}.

\begin{table}[tb]
\centering
    \caption{\label{tbl:real}Estimates of the fit for power-law scaling of network density and size in $50$ real-world networks revealed under different renormalization techniques. Values are estimates of the mean over $10$ renormalizations of each network and correspond to correlation coefficient~$\rho$, coefficient of determination~$R^2$, expressed network density~$d$ and the number of revealed self-similar scales~$S$. (For each technique, $\rho$~and~$R^2$ are obtained separately for original and renormalized networks, and for renormalized varieties only---first and second row, respectively. Bold values of $R^2$ indicate relatively high goodness of fit to a power-law, whereas values in italics show poor performance of the respective renormalization technique.)}
   \begin{tabular}{ccccc}
   \\\hline\noalign{\smallskip}
   Technique & $\rho$ & $R^2$ & $d$ & $S$\\\noalign{\smallskip}\hline\noalign{\smallskip}
	\multirow{2}{*}{Randomized box-tiling} & $-0.975$ & $\mathbf{0.944}$ & \multirow{2}{*}{$1.7\cdot n^{-0.807}$} & \multirow{2}{*}{$5.3$}\\
	 & $-0.973$ & $\mathbf{0.936}$ & \\\noalign{\smallskip}
	\multirow{2}{*}{Randomized cluster-growing} & $-0.977$ & $\mathbf{0.948}$ & \multirow{2}{*}{$1.6\cdot n^{-0.818}$} & \multirow{2}{*}{$4.6$}\\
	 & $-0.977$ & $\mathbf{0.944}$ & \\\noalign{\smallskip}
	\multirow{2}{*}{Balanced propagation} & $-0.985$ & $\mathbf{0.962}$ & \multirow{2}{*}{$1.9\cdot n^{-0.836}$} & \multirow{2}{*}{$4.3$}\\
	 & $-0.980$ & $\mathbf{0.963}$ & \\\noalign{\smallskip}
	\multirow{2}{*}{Modularity optimization} & $-0.966$ & $\mathbf{0.956}$ & \multirow{2}{*}{$3.0\cdot n^{-0.882}$} & \multirow{2}{*}{$3.9$}\\
	 & $-0.889$ & $\mathit{0.820}$ & \\\noalign{\smallskip}
	\multirow{2}{*}{Spectral analysis} & $-0.951$ & $0.922$ & \multirow{2}{*}{$4.1\cdot n^{-0.893}$} & \multirow{2}{*}{$4.5$}\\
	 & $-0.878$ & $\mathit{0.718}$ & \\\noalign{\smallskip}\hline\noalign{\smallskip}
	Original networks & $-0.924$ & $0.870$ & $3.8\cdot n^{-0.921}$ & \\\noalign{\smallskip}\hline
    \end{tabular}
\end{table}

\begin{figure}[p]
\centering
\includegraphics[width=1.00\columnwidth]{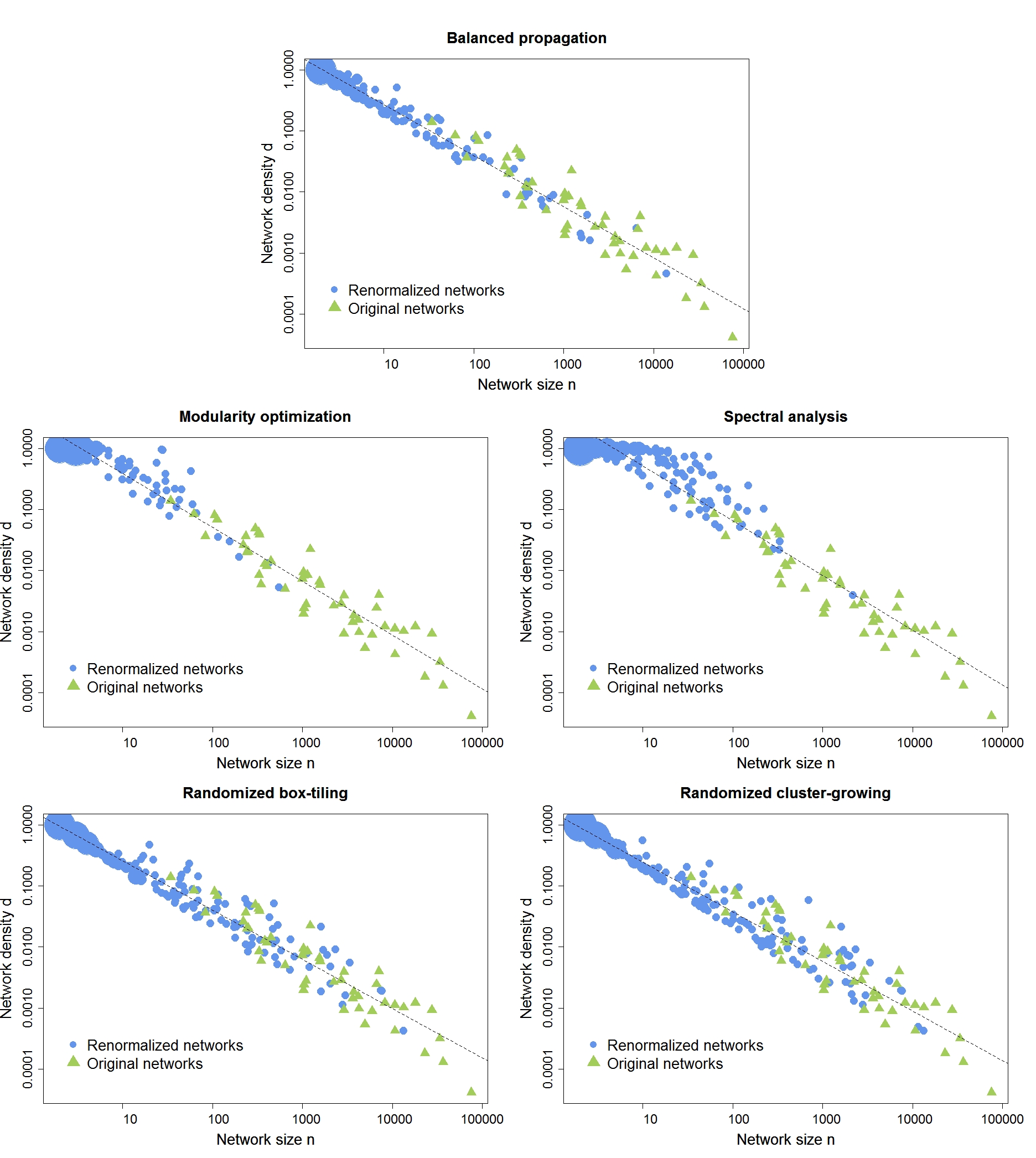}
\caption{\label{fig:real}Power-law scaling of network density and size in $50$ real-world networks of diverse origin revealed with different renormalization techniques. Plots show scaling of density for a single renormalization of each network under respective technique. (Green triangles correspond to original networks, whereas blue circles represent their renormalized varieties. Symbol sizes are proportional to the number of networks with the same size and density.)}
\end{figure}

The plots on \figref{real} illustrate size and density relationships with the scaling exponents $\gamma$ around $-0.85$. Original networks exhibit greater scaling factor (see also \secref{analys:large}), which indicates $\gamma$ is approaching $-1$ for adequately large $n$. This corresponds to commonly observed finding that most large-scale real-world networks tend to be sparse---the number of links appears not to be close to $O(n^2)$ but rather of order $O(n)$. Consecutively, we can simplify density definition with the relationship $d \approx n^{-1}$. Thus, power-law relationship between the network size and density is expected for original networks (without considering reduced varieties). However, among renormalized networks the relationships follow even stronger power-laws. This means that networks obtained on different scales of renormalization process also satisfy power-law relationship between size and density, and implies an existence of density scaling also within real-world networks.

Furthermore, results show similar behavior of exponents $\gamma$ and constants $c$ for better performing techniques, including box-tiling, cluster-growing, and balanced propagation. This finding implies that box-covering methods find smaller and sparser boxes, similar to communities detected with balanced propagation. Other two algorithms reveal bigger, denser, and also more heterogeneous communities considering density scaling. The values of self-similar scales $S$ are in accordance with these observations. Modularity optimization extracts network with one community in the least number of scales on average (bigger communities). On the other hand, box-tiling obtains a larger number of reduced networks (smaller boxes), which is expected due to the distance $l_B$ setting.

To summarize, the analysis of real-world networks reveals power-law scaling of the network density with respect to network size. Among the employed renormalization techniques, balanced propagation seems to lead to the most optimal reduction of networks according to the density scaling. Results acquired by three best performing techniques indicate an existence of a certain common organizing principle of networks, which dictates linking rules and interactions among nodes. Our findings thus advance a recent discovery of a universal scaling of density among real-world networks~\cite{LJTBH11}, since we reveal density scaling also among different self-similar scales of complex real-world networks. In addition, the results are consistent with the densification laws of Leskovec et al.~\cite{LKF07, LKF05}---$m \propto n^\alpha$, where $\alpha$ ranges between $1$ and $2$ and relates with our exponent $\gamma$, which lies between $0$ and $-1$ respectively. Thus, our study expands densification laws to other dimensions of network structure.

Besides density, we also studied the scaling of other network properties with respect to network size. In particular, we analyzed number of links, average and maximum degree, number of articulation points, average path and diameter~\cite{WS98}, betweenness and closeness centrality~\cite{Fre79} and clustering coefficient~\cite{Was94}. The results reveal significant scaling also between network size and average node or link betweenness---the number of shortest paths going through a node and link respectively. Regarding to a definition of network density and observed power-law relationship between size and density, similar relationship for number of links occurs expectedly. However, due to simplicity, detailed investigation of betweenness centrality scaling is omitted, although a prominent direction for future research.


\subsection{\label{sec:analys:rand}Random networks}

To further validate our results we apply box-tiling and modularity optimization renormalizations to Erd\H{o}s-R{\'e}nyi random graphs with different sizes $n$ and probabilities of linking nodes $p$. We generate networks with $500$, $1000$, $2500$, $5000$, and $10000$ nodes and probabilities corresponding to density obtained with modularity optimization based renormalization (\secref{analys:real}), density reported in~\cite{LJTBH11}, and probability that should assure sufficient size of the largest network component~\cite{ER59}.

Firstly, we test balanced propagation renormalization, since the method performs best on real-world networks. The results prove to be very good, showing fits closely to ideal ($R^2$ and $\rho$ close to $1$ and $-1$, respectively). However, detailed investigation shows renormalization for most of the generated networks reveals only a single scale or concludes without reduction, since random networks supposedly have no community structure. For this reason we exclude balanced propagation from the analysis. Thus, we study box-tiling as an illustration of classical box-covering principle and modularity optimization as an example of community based renormalization. Note that, in contrast to the above, the latter reveals non-trivial modules also in random networks (\secref{analys:real}).

The results appear in \tblref{rand}. A strong relationship ($R^2=1$, $\rho=-1$) arises between size and density of original networks. That occurs due to the settings of probability $p$. These strong fits cause also high values for original and randomized networks together. The results for randomized varieties of networks show low fits to the data and implies rather diverse density of reduced networks with respect to their size. This is anticipated owing to random network structure. However, the values of $R^2$ and $\rho$ for randomized networks under $p=2/(n-1)$ setting are relatively high. Examining plot for box-tiling closely shows diverse density among reduced networks, however, diversity straightens due to the large number of reduction scales. On the other hand, networks reduced under modularity optimization on each scale reveal almost the same density, and thus lead to higher fit. Slightly greater values for renormalized networks under box-tiling seem to occur due to the definition of boxes, which consider only proximity among nodes.

\begin{table}[tb]
\centering
    \caption{\label{tbl:rand}
Estimates of the fit for power-law scaling of network density and size in Erd\H{o}s-R{\'e}nyi~random graphs obtained with two renormalization techniques. For each probability of a link between two nodes~$p$, we construct an ensemble of networks of various sizes. Values are estimates of the mean over $10$ realizations of each random graph. (See also \tblref{real}.)}
   \begin{tabular}{cccccc}
   \\\hline\noalign{\smallskip}
   $p$ & Technique & $\rho$ & $R^2$ & $d$ & $S$ \\\noalign{\smallskip}\hline\noalign{\smallskip}
	\multirow{4}{*}{$3.0\cdot n^{-0.882}$} & \multirow{2}{*}{Randomized box-tiling} & $-0.925$ & $\mathit{0.820}$ & \multirow{2}{*}{$2.8\cdot n^{-0.753}$} & \multirow{2}{*}{$4.9$} \\
	& & $-0.852$ & $\mathit{0.787}$ \\\noalign{\smallskip}
	& \multirow{2}{*}{Modularity optimization} & $-0.957$ & $\mathbf{0.994}$ & \multirow{2}{*}{$12.3\cdot n^{-0.963}$} & \multirow{2}{*}{$3.0$} \\
	& & $-0.583$ & $\mathit{0.446}$\\\noalign{\smallskip}\hline\noalign{\smallskip}
	\multirow{4}{*}{$7.9\cdot n^{-0.986}$} & \multirow{2}{*}{Randomized box-tiling} & $-0.939$ & $\mathit{0.818}$ & \multirow{2}{*}{$3.7\cdot n^{-0.797}$} & \multirow{2}{*}{$4.5$} \\
	& & $-0.882$ & $\mathit{0.781}$ \\\noalign{\smallskip}
	& \multirow{2}{*}{Modularity optimization} & $-0.964$ & $\mathbf{0.998}$ & \multirow{2}{*}{$10.5\cdot n^{-1.022}$} & \multirow{2}{*}{$3.0$} \\
	& & $-0.494$ & $\mathit{0.537}$ \\\noalign{\smallskip}\hline\noalign{\smallskip}
	\multirow{4}{*}{$2/(n-1)$} & \multirow{2}{*}{Randomized box-tiling} & $-0.990$ & $\mathbf{0.967}$ & \multirow{2}{*}{$2.6\cdot n^{-0.953}$} & \multirow{2}{*}{$6.7$} \\
	& & $-0.986$ & $\mathbf{0.962}$ \\\noalign{\smallskip}
	& \multirow{2}{*}{Modularity optimization} & $-0.930$ & $0.916$ & \multirow{2}{*}{$6.4\cdot n^{-1.065}$} & \multirow{2}{*}{$4.0$} \\
	& & $-0.744$ & $\mathit{0.817}$\\\noalign{\smallskip}\hline
    \end{tabular}
\end{table}

Other variables, including scaling exponent $\gamma$, constant $c$, and revealed self-similar scales $S$, comprehend greater range than values for real-world networks. This verifies there exists no optimal density characteristic for random networks and denotes that random networks do not exhibit common power-law density scaling.

According to the above, we conclude that results for random networks appear to be weak as anticipated, since random networks should not reveal structures like communities in real-world networks. On the contrary, findings indicate that self-similar density scaling of real-world networks is not obtained by chance, and the scaling exists due to some inner principles which determine network structure.

We have also analyzed several generative graph models, whether they reveal similar scaling of density as observed in real-world networks. Expectedly, under balanced propagation renormalization, classical scale-free~\cite{BA99,EW02} and small-world graph~\cite{Kle00} models show the same behavior as in the case of random graphs (due to the lack of community structure). On the other hand, forest fire model proposed by Leskovec~et~al.~\cite{LKF07,LKF05} reveals strong power-law scaling of density with estimates similar to those observed in~\tblref{real} (exact results are omitted). Interestingly, the model gives networks that also obey network densification laws, shrinking diameters, community structure and scale-free degree distributions~\cite{LKF07}, and thus provide a relatively realistic structure of real-world networks. Note that community guided attachment model~\cite{LKF07,LKF05} that also follows densification laws, does not reveal self-similar scaling of density; thus, the latter is indeed not an artifact of the former, but rather extends network density laws to other dimensions.


\subsection{\label{sec:analys:large}Large real-world networks}

For a complete analysis, we also analyze the size and density relationship of the largest five real-world networks presented in \tblref{nets}. In particular, co-purchase network of different products from Amazon in 2006, complete map of \textit{nd.edu} domain, road network of Pennsylvania, communication network of user discussions on Wikipedia before January 2008, and Internet topology graph from traceroutes in 2005. Due to simplicity, we present study only for the best performing balanced propagation based renormalization, where the maximum number of iterations is limited to 100.

\begin{figure}[tbh]
\centering
\includegraphics[width=1.00\columnwidth]{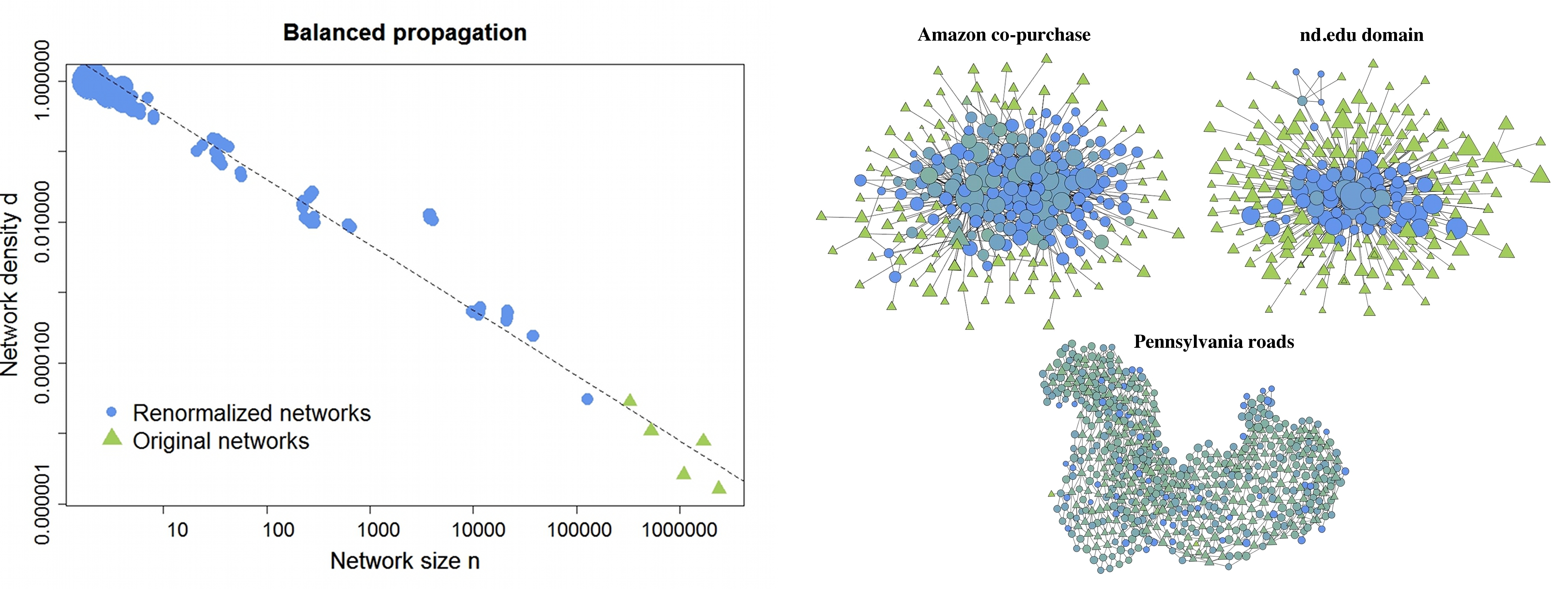}
\caption{\label{fig:large}(left)~Power-law scaling of network density and size in five real-world networks with millions of links revealed with balanced propagation. Plot shows scaling of density over $10$ renormalizations of each network. (Green triangles correspond to original networks, whereas blue circles represent their renormalized varieties. Symbol sizes are proportional to the number of networks with the same size and density.) (right)~Density of network structure in renormalized varieties of three large real-world systems of different origin. (Node symbols correspond to degree-corrected clustering coefficient~\cite{SV05} that ranges between $0$ and $1$---green triangles and blue circles, respectively---while symbol sizes are proportional to the number of nodes in the original network.)}
\end{figure}

The results are presented in \tblref{large}. Observing only original networks, fits are expectedly low due to small number of networks considered. For the same reason the constant $c$ and exponent $\gamma$ also differ from the ones in \secref{analys:real}. However, other results show very good fit particularly for original and randomized networks together and reveal a power-law relationship of network size and density (see \figref{large}). (Again, the results are statistically significant at one percent significance level.) As expected due to the size of the networks, the scaling exponent is close to $-1$. Number of self-similar scales is higher as in analysis in \secref{analys:real}, since networks are larger and thus reduced in more steps. On the other hand, $S$ does not significantly increase with network size, which implies that renormalization is effective and efficient approach for simplifying large networks.

\begin{table}[tb]
\centering
    \caption{\label{tbl:large}Estimates of the fit for power-law scaling of network density and size in five large real-world networks revealed with balanced propagation. Values are estimates of the mean over $10$ renormalizations of each network. (See also \tblref{real}.)}
   \begin{tabular}{ccccc}
   \\\hline\noalign{\smallskip}
   Technique & $\rho$ & $R^2$ & $d$ & $S$\\\noalign{\smallskip}\hline\noalign{\smallskip}
	\multirow{2}{*}{Balanced propagation} & $-0.990$ & $\mathbf{0.977}$ & \multirow{2}{*}{$2.9\cdot n^{-0.926}$} & \multirow{2}{*}{$4.9$}\\
	 & $-0.980$ & $\mathbf{0.961}$ & \\\noalign{\smallskip}\hline\noalign{\smallskip}
	Original networks & $-0.900$ & $\mathit{0.719}$ & $66.2\cdot n^{-1.175}$ & \\\noalign{\smallskip}\hline
    \end{tabular}
\end{table}

\figref{large} illustrates renormalized varieties of three large networks. We consider networks of diverse origin to value how different structure of networks effects the relationship between size and density. For instance, Pennsylvania roads network shows very homogeneous structure, while, on the contrary, other two networks present core-periphery structure typical for social and information networks. However, these diverse network structures do not reflect in the results (\figref{large}, left). Thus, the finding confirms common density scaling in real-world networks irrespective of network type and origin.

Our study improves the comprehension of self-similar structure in real-world networks and implies several possible applications. Firstly, adequate network coarse-graining implies simplification and abstraction of large real-world networks without losing information about original network density. Reduction also enables visualization and improves the comprehension of larger complex networks. Additionally, self-similar density scaling can help at detecting sufficient density according to the size of the sub-graphs in graph sampling applications (e.g.,~\cite{LF06}), improve the accuracy of link prediction (for review see~\cite{LZ11}) and the quality of synthetic graph generation (e.g.,~\cite{HP11}).


\section{\label{sec:conc}Conclusions}

The paper explores the relationship between size and density of complex real-world networks under different box-covering and community-based renormalization techniques. The analysis was conducted on over 50 real-world networks of various sizes as well as Erd\H{o}s-R{\'e}nyi random graphs and different generative graph models. The main contribution of the study is to imply an existence of a scale-free density not only among different real-world networks, but also among their self-similar scales. Common scaling of density thus appears to be a unique property of complex real-world networks irrespective of their type, size and origin. Also, the results reveal balanced propagation based renormalization as the best performing method among the observed algorithms. The study on Erd\H{o}s-R{\'e}nyi random graphs, which supposedly exhibit no community structure, validates the above results and confirms that observed scaling of density is distinctive for real-world networks. Hence, our findings expand recent discoveries to other dimensions of network structure and further improve the comprehension of self-similarity in complex real-world networks. The latter has possible applications in graph sampling, link prediction, synthetic graph generation, network abstraction and visualization.

In our future work we intend to focus on other possible characteristics of density scaling, that could be identified in networks of common type and origin. Furthermore, we will analyze the betweenness centrality scaling with respect to network size in detail. Moreover, the work will also be extended on finding suitable ways for abstracting large real-world networks, while at the same time preserving their fundamental properties.


\section*{Acknowledgment}
Authors would like to thank Matija Polajnar for providing Slovenian scientists collaboration data and Bojan Klemenc for useful comments. The work has been supported by the Slovene Research Agency \textit{ARRS} within the research program P2-0359.


\bibliographystyle{elsarticle-num}


\end{document}